\documentclass[twocolumn,american,english,aps]{revtex4}
\usepackage[T1]{fontenc}
\usepackage[latin1]{inputenc}
\usepackage{graphicx}
\usepackage{amssymb}

\makeatletter

\makeatletter



\makeatother

\makeatother

\usepackage{babel}

\begin{document}

\title{Hilbert Space Average Method and adiabatic quantum search }

\author{A. Pérez}

\affiliation{Departament de Física Teòrica and IFIC, Universitat de València-CSIC
\\
 Dr. Moliner 50, 46100-Burjassot, Spain\\
}

\begin{abstract}
We discuss some aspects related to the so-called Hilbert space Average
Method, as an alternative to describe the dynamics of open quantum
systems. First we present a derivation of the method which does not
make use of the algebra satisfied by the operators involved in the
dynamics, and extend the method to systems subject to a Hamiltonian
that changes with time. Next we examine the performance of the adiabatic
quantum search algorithm with a particular model for the environment.
We relate our results to the criteria discussed in the literature
for the validity of the above-mentioned method for similar environments.
\end{abstract}
\maketitle

\section{Introduction}

The study of open quantum systems has attracted renewed attention
during the last years. One important reason for this is the expected
advent of future quantum computers \cite{Chuang}. The interaction
of the quantum computer with its surroundings can introduce some degree
of decoherence which can, eventually, ruin the performance of the
quantum algorithm. Although this phenomenon can be mitigated with
the help of error-correction methods, a deeper understanding of how
the ambiance operates on the smaller system can also be used to improve
the working conditions. In fact, recent papers have shown that, for
some models of system-ambiance interaction, the loss of coherence
can be smaller if the ambiance temperature is increased \cite{Amin,Montina}.
Similarly, designing some engineered reservoirs with controlled coupling
and state of the environment can reduce the decoherence rate \cite{Cui,Gordon}.
One can even consider purely dissipative processes, which turn out
to be equivalent to a quantum circuit model for quantum computation
\cite{Verstraete}. On the other hand, systems subject to decoherence
will experience a transition from a quantum to a classical state.
The study of this transition will give more insight about the nature
of Quantum Mechanics and its differences with a classical perception
\cite{Ollivier}.

There are different approaches which have been developed in the literature
in order to describe the evolution of open systems, based on different
techniques such as master equations or superoperators \cite{Breuer}.
As an alternative to these methods, we will study the behavior of
the open system using the so-called Hilbert space Average Method (HAM,
in what follows) \cite{Michel,Gemmer}. This method has been proved
to give, in some situations, better results than conventional Time
Convolutionless (TCL) approximations, and comparable to correlated
projection superoperator techniques \cite{Breuer06}. We will extend
this approach to the case of a time-varying Hamiltonian acting on
the open system, and will show that, under a suitable choice of the
operators defining the HAM scheme, the resulting equations will adopt
the form of traditional master equations, at least up to second order
in the system-environment coupling.

As an application, we will consider the case of a quantum system which
is designed to perform a Grover search \cite{Grover,Boyer,Grover2}
via adiabatic quantum computation \cite{Farhi2,Farhi3,Roland,Das}.
Our purpose is to analyze the response of the quantum computer when
coupled to the external influence of an environment, introduced with
the help of some specific model. This problem has been considered
by several authors \cite{Amin,Childs,Sarandy,Ashhab,Amin08,Tiersch,Perez07,Aabergrc,Aaberg,Johansson},
but here we relate it to similar models for the bath which have discussed
within the HAM formalism. We compare our results with a numerical
simulation of the Schrödinger equation obeyed by the full system. 

In Section II we introduce some basic notations. In Sect. III we briefly
revisit the HAM method, using an approach that makes not use of the
algebra of the involved operators. An approximated evolution equation,
extended to the case of a time-changing Hamiltonian acting on the
system, is obtained in Sect. IV, and we also make a connection with
familiar master equations. Sect. V is devoted to the analysis of adiabatic
search when the quantum computer interacts with a particular environment.
The evolution of the system is followed both by the exact Schrödinger
equation and by solving the obtained approximated equations. The comparison
of both calculations is discussed within the framework of known criteria
for similar models, which have derived within the HAM formalism. Our
results are summarized on Sect. VI.

We work in units such that $\hbar=1$.

\section{basic notations}

We wish to study the evolution of an open system (S) in contact with
an environment (E). First we introduce the basic quantities in the
Schrödinger picture, and then we will define a convenient interaction
picture for this problem. System S is subject to a time-dependent
Hamiltonian $H_{S}(t)$ . Let us denote by $H_{E}$ the Hamiltonian
describing the free evolution of the environment, and by $V$ the
interaction between both systems. Both $H_{E}$ and $V$ are assumed
to be time-independent. Moreover, we make the hypothesis that $[H_{S}(t),H_{E}]=0$
. The evolution of the density matrix $\rho(t)$ of the complete (S+E)
system is therefore given by\begin{equation}
\frac{d}{dt}\rho(t)=-i[H(t),\rho(t)]\end{equation}

with $H(t)=H_{S}(t)+H_{E}+V$ the total Hamiltonian. We can define
an interaction picture density matrix $\rho_{I}(t)$ as follows:\begin{equation}
\rho_{I}(t)\equiv\exp(iH_{E}t)\rho(t)\exp(-iH_{E}t).\end{equation}

The equation for $\rho_{I}(t)$ is easily obtained:\begin{equation}
\frac{d}{dt}\rho_{I}(t)=-i[H_{S}(t),\rho_{I}(t)]-i[V_{I}(t),\rho_{I}(t)],\label{exactdynamic}\end{equation}

where \begin{equation}
V_{I}(t)\equiv\exp(iH_{E}t)V\exp(-iH_{E}t)\end{equation}

is the interaction operator in the interaction picture. In what follows
we will assume, unless otherwise specified, that we are working in
the above-defined picture, and will therefore omit the subscript 'I'.

\section{Hilbert Space Average method }

In this section we briefly review the Hilbert space Average Method
(HAM), as an alternative to describe the dynamics of an open quantum
system. A more detailed description of the method can be found in
\cite{Michel,Breuer06,Gemmer}. 

The idea is to replace the dynamics of the density matrix $\rho(t)$
describing the full system (i.e. open system plus reservoir) by an
\textit{effective} density matrix $\alpha(t)$ which is simpler to
describe, with the condition that the expected values of a given set
of operators is reproduced. Let us assume that we are interested on
a set of operators $\{\hat{P}_{n}\}$ and we want to define a density
matrix satisfying the boundary conditions\begin{equation}
Tr[\alpha(t)\hat{P}_{n}]\equiv p_{n}(t),\label{conditions}\end{equation}

where the functions $p_{n}(t)$ are assumed to be known (actually,
they will be determined by the dynamics), and $Tr$ stands for the
trace over the whole Hilbert space. We would like to determine $\alpha(t)$
respecting the above conditions, but otherwise unknown. To this end
we establish the following procedure. We maximize the entropy, in
order to account for our ignorance about the effective density matrix,
but add constraints corresponding to Eqs. (\ref{conditions}) under
the form of Lagrange multipliers. For our purpose, it is simpler to
consider the linear entropy $S[\rho]=1-Tr[\rho^{2}]$. In this way,
we find the extrema of the functional\begin{equation}
I[\alpha]\equiv S[\alpha]-\sum_{n}a_{n}Tr[\alpha\hat{P}_{n}]\end{equation}

with $\{a_{n}\}$ the above defined Lagrange multipliers. Following
this procedure, one arrives to the expression\begin{equation}
\alpha(t)=\sum_{n}b_{n}(t)\hat{P}_{n},\label{alpha}\end{equation}

with $b_{n}=-a_{n}/2$ , and we have made explicit the time dependence
of $\alpha(t)$ . The new functions $b_{n}(t)$ are determined by
Eqs (\ref{conditions}). Of course, one also has to make sure that
the condition $Tr[\alpha(t)]=1$ is satisfied.

In \cite{Michel,Breuer06,Gemmer}, the authors introduce the HAM
method as an average over all possible states in the Hilbert space
that accounts for conditions Eqs. (\ref{conditions}) (here is where
the name HAM comes from), and introduce the set of operators $\{\hat{P}_{n}\}$
as obeying a closed algebra. In contrast, our derivation of Eq. (\ref{alpha}),
although perhaps less intuitive, makes no use of the algebra of the
operators $\{\hat{P}_{n}\}$ .

\section{Approximated evolution equation}

The exact dynamics of $\alpha(t)$ will be given by solving an equation
like Eq. (\ref{exactdynamic}) in the interaction picture. This equation
does not admit a simple, closed form, solution. In this section we
will investigate some approximation that is easier to solve, and will
allow us to make a connection with standard master equations. We first
introduce the evolution operator $U(t+\tau,t)$ from instant $t$
to $t+\tau$. We then have\begin{equation}
\alpha(t+\tau)=U(t+\tau,t)\alpha(t)U^{\dagger}(t+\tau,t).\label{evolalpha}\end{equation}

In order to separate the evolution due to $H_{s}(t)$ from that due
to $V(t)$ we consider a sufficiently small $\tau$ and approximate
$U(t+\tau,t)$ by a second-order Suzuki decomposition \cite{Suzuki}\begin{equation}
U(t+\tau,t)\simeq\exp(-i\frac{\tau}{2}H_{S}(t))D_{V}(t+\tau,t)\exp(-i\frac{\tau}{2}H_{S}(t)),\label{Suzuki}\end{equation}

where $D_{V}(t+\tau,t)$ is the evolution operator describing the
time evolution due to $V(t)$ alone (i.e., neglecting $H_{S}(t)$
in the Hamiltonian), and verifies the equation \begin{equation}
i\frac{d}{dt}D_{V}(t,t_{0})=V(t)D_{V}(t,t_{0})\label{eqDV}\end{equation}
Inserting Eq. (\ref{Suzuki}) into (\ref{evolalpha}) and expanding
the exponentials in powers of $\tau$ gives \begin{equation}
\alpha(t+\tau)=\alpha(t)+\Delta\alpha(t+\tau,t)+i\tau[\alpha(t),H_{S}(t)]+\mathcal{O}(\tau^{2}),\end{equation}

with the definition \begin{equation}
\Delta\alpha(t+\tau,t)\equiv D_{V}(t+\tau,t)\alpha(t)D_{V}^{\dagger}(t+\tau,t)-\alpha(t).\label{defdeltaalpha}\end{equation}

We now would like to make an approximate treatment of the quantity
$\Delta\alpha(t+\tau,t)$ defined in the latter equation. To this
purpose, we use the solution of Eq. (\ref{eqDV}) up to second order
in the potential\begin{equation}
D_{V}(t+\tau,t)\simeq I-i\int_{t}^{t+\tau}dsV(s)-\int_{t}^{t+\tau}ds\int_{t}^{s}ds'V(s)V(s').\end{equation}

Within this approximation, Eq. (\ref{defdeltaalpha}) reads

\begin{eqnarray}
\Delta\alpha(t+\tau,t) & = & -i\int_{t}^{t+\tau}ds[V(s),\mbox{}\alpha(t)]\nonumber \\
 &  & \hspace{-1cm}-\int_{t}^{t+\tau}ds\int_{t}^{s}ds'[V(s),[V(s'),\alpha(t)]]\,.\label{aproxdeltaalpha}\end{eqnarray}
Starting from this approximation, one can derive the corresponding
equations for the functions $\{p_{n}(t)\}$ , following the procedure
described in \cite{Michel}. To this end, one needs to specify the
operators $\{\hat{P}_{n}\}$ and the algebra verified by them. 

We can also establish a connection with familiar master equations,
which is done in a trivial way within the above formalism. We simply
assume that the effective density matrix $\alpha(t)$ can be factorized
as\begin{equation}
\alpha(t)=\rho_{S}(t)\otimes\rho_{E}\label{alphasep}\end{equation}

where $\rho_{E}$ is a density matrix that approximates the state
of the bath, and $\rho_{S}(t)$ is the density matrix for the system
S, related to $\alpha(t)$ via \begin{equation}
\rho_{S}(t)=Tr_{E}[\alpha(t)],\end{equation}

and $Tr_{E}$ indicates the partial trace over the environment E. 

Let us introduce an orthonormal basis $\{|i>\}$ in the Hilbert space
corresponding to system S, and write Eq. (\ref{alphasep}) in the
following way:\begin{equation}
\alpha(t)\equiv\sum_{i,j}P_{ji}(t)|i><j|\otimes\rho_{E}\end{equation}

By comparing with Eq. (\ref{alpha}) we identify the operators $\hat{P}_{n}$
associated with the ansatz (\ref{alphasep})\begin{equation}
\hat{P}_{n}=|i><j|\otimes\rho_{E}\end{equation}

where $n$ indicates a given pair $i,j$. One also easily obtains
from Eq. (\ref{conditions}) that\begin{equation}
p_{n}(t)=P_{ij}(t)Tr[\rho_{E}^{2}].\end{equation}

Notice that the functions $P_{ij}(t)$ are related to the matrix elements
$\rho_{Si,j}(t)$ in the basis $\{|i>\}$ via $P_{ij}(t)=\rho_{Sji}(t)$
. 

In order to obtain a more detailed expression for the quantity $\Delta\alpha(t+\tau,t)$
one needs to specify the interaction $V$ . Let us assume that this
operator is defined, in the Schrödinger picture, by \begin{equation}
V=\sum_{i}A_{i}\otimes C_{i}\label{VprodS}\end{equation}

with the Hermitian operators $A_{i}$ ($C_{i}$) acting on the system
S (E). Of course, one can consider a more general situation where
these operators are not Hermitian, and simply add the Hermitian conjugate
to $V$ . However, a simplified version like Eq. (\ref{VprodS}) will
be sufficient to our purposes. In the interaction picture, the above
formula becomes\begin{equation}
V_{I}(t)=\sum_{i}A_{i}\otimes C_{i}(t)\label{VprodI}\end{equation}

with $C_{i}(t)\equiv\exp(iH_{E}t)C_{i}\exp(-iH_{E}t)$ . Hereafter,
we omit the subindex 'I', as anticipated in Sect. II, and assume that
we are working in the interaction picture.

In what follows, we are interested in the difference\begin{equation}
\Delta\rho_{S}(t+\tau,t)\equiv\rho_{S}(t+\tau)-\rho_{S}(t)=Tr_{E}[\Delta\alpha(t+\tau,t)].\end{equation}

We will obtain an approximation to this quantity by using Eqs. (\ref{aproxdeltaalpha})
and (\ref{VprodI}). The rest of this section is a standard manipulation
which is common to the derivation of master equations. Our purpose
is only to show that the program we started in Sect. III does indeed
lead to such kind of equations. The interested reader is addressed
to the existing bibliography (see, e.g. \cite{Breuer}). The final
expression reads\begin{eqnarray}
\frac{d}{dt}\rho_{S}(t) & = & -i[H_{S}(t),\rho_{S}(t)]\nonumber \\
 & \hspace{-2.8cm}+ & \hspace{-1.5cm}\frac{1}{2}\sum_{l,k}\Gamma_{lk}(t)\{A_{k}\rho_{S}(t)A_{l}-A_{l}A_{k}\rho_{S}(t)\}+\mathrm{h.c.}\label{mastereq}\end{eqnarray}

where the Hermitian conjugate refers only to the summation. In obtaining
Eq. (\ref{mastereq}) we have taken the limit $\tau\rightarrow0$
, and we have defined \begin{equation}
\Gamma_{lk}(t)=\lim_{\tau\rightarrow0}\frac{2}{\tau}\int_{t}^{t+\tau}ds\int_{t}^{s}ds'G_{lk}(s,s'),\label{gammalk}\end{equation}

with \begin{equation}
G_{lk}(s,s')\equiv Tr_{E}\{C_{l}(s)C_{k}(s')\rho_{E}\}\label{correlations}\end{equation}

the bath correlation functions. We also have made the usual hypothesis
\cite{Breuer} that \begin{equation}
Tr_{E}\{C_{l}(s)\rho_{E}\}=0.\label{trazacero}\end{equation}

Eq. (\ref{mastereq}) takes then the familiar form of a master equation,
which becomes of the Lindblad type in the case that the coefficients
$\Gamma_{lk}(t)$ are independent of time.

\section{Model for adiabatic search and interaction with the environment.}

We now analyze a particular and interesting example, which can be
cast either under the form of HAM or master equations, according to
the discussion of the previous section. We study the performance of
an open quantum system (the quantum computer) consisting on $n$ qubits,
while it does an adiabatic search for a marked state $|m>$ out of
$N=2^{n}$ possible configurations, subject to the interaction with
an environment. This problem has been addressed by several authors
\cite{Amin,Childs,Sarandy,Ashhab,Amin08,Tiersch,Perez07,Aabergrc,Aaberg,Johansson}.
Here, we relate the problem to similar bath models which have derived
within the HAM formalism.

The Hamiltonian $H_{S}(t)$ implements the adiabatic quantum search,
and will be written as 

\begin{equation}
H_{S}(t)=f(t)(I-|\Psi_{0}><\Psi_{0}|)+g(t)(I-|m><m|)\label{Hs(t)}\end{equation}

where $|\Psi_{0}>$ corresponds to the initial state of the system,
which we take as the equally-weighted superposition $|\Psi_{0}>=\frac{1}{\sqrt{N}}\sum_{i=1}^{N}|i>$
and $I$ is the identity operator. The functions $f(t)$ and $g(t)$
will vary slowly during the running time $t_{G}$, and satisfy $f(0)=1$,
$g(0)=0$ , $f(t_{G})=0$, $g(t_{G})=1$. There are many possible
choices of these functions, depending on the trade-off between time
and energy cost one pursues \cite{Roland,Das,Perez07}. Here we choose
these functions as obtained form imposing a local adiabatic condition
\cite{Roland}, with $f(t)=1-s(t)$ , $g(t)=s(t)$ . In the large
$N$ limit 

\begin{equation}
s(t)=\frac{1}{2}+\frac{1}{2\sqrt{N}}\tan(\frac{2\epsilon t}{\sqrt{N}}-\arctan\sqrt{N}),\end{equation}

where $\epsilon$ is a small number that controls the probability
of success for the algorithm, which will run during a time $t_{G}=\pi\sqrt{N}/2\epsilon$. 

The adiabatic quantum search evolution can be effectively reduced
to a two-level system, in the space spanned by the orthogonal vectors
$\{|m>,|p>\}$ , with $|p>=\frac{1}{\sqrt{N-1}}\sum_{i\neq m}|i>$.
The minimum energy gap in the Hamiltonian (\ref{Hs(t)}) appears for
eigenstates which are linear combinations of $\{|m>,|p>\}$ . It occurs
when $s(t)\simeq1/2$ and takes the value $1/\sqrt{N}$. The rest
of eigenvectors are degenerate, with eigenvalue $f+g=1$, and are
well separated from the previous two, specially around the avoided
crossing point $s\simeq1/2$. It is reasonable to assume that one
can restrict the system evolution to this effective two-level space.
Arguments to favour this assumption are shown in \cite{Amin08}.

\selectlanguage{american}%
We now introduce a model to describe the interaction with the environment.
We will make an explicit comparison with a numerical simulation of
the \foreignlanguage{english}{Schrödinger equation. To this end, we
make use of a simple model consisting on a band of $N_{1}$ equally
spaced levels. As we show, this model can describe relaxation to equilibrium
and decoherence effects in a natural way, and may be regarded as a
simplified version of the two-band model described in \cite{Breuer06}
. The Hamiltonian describing the environment is given by} \begin{eqnarray}
H_{E} & = & \sum_{n=1}^{N_{1}}\frac{\delta\varepsilon}{N_{1}}n|n\rangle\langle n|\end{eqnarray}
 and the interaction between both systems by \begin{equation}
V=\sum_{i=1}^{n}\sigma_{+}^{i}B_{i}+{\mbox{h.c.}},\label{MODEL-V}\end{equation}

with \begin{equation}
B_{i}=\lambda_{i}\sum_{n_{2}>n_{1}}c_{i}(n_{1},n_{2})|n_{1}\rangle\langle n_{2}|.\label{couplings}\end{equation}

The indices $n,n_{1}$ and $n_{2}$ label the levels of the energy
band, and $\sigma_{+}^{i}$ are Pauli matrices acting on qubit $i$
. The global strength of the interaction with each one of the qubits
is given by $\lambda_{i}$. The coupling constants $c_{i}(n_{1},n_{2})$
are independent Gaussian random variables. In order to make the model
simpler, we will choose the same couplings for all qubits, which amounts
to the replacement $\sum_{i=1}^{n}\lambda_{i}c_{i}(n_{1},n_{2})\rightarrow n\lambda c(n_{1},n_{2})$
in Eq. (\ref{MODEL-V}). The averages (denoted by $<>$) over the
random constants $c(n_{1},n_{2})$ satisfy: \begin{eqnarray}
\langle c(n_{1},n_{2})\rangle & = & 0,\label{AV1}\\
\langle c(n_{1},n_{2})c(n'_{1},n'_{2})\rangle & = & 0,\label{AV2}\\
\langle c(n_{1},n_{2})c^{*}(n'_{1},n'_{2})\rangle & = & \delta_{n_{1},n'_{1}}\delta_{n_{2},n'_{2}}.\label{AV3}\end{eqnarray}

\selectlanguage{english}%
We will assume that an average over the possible realizations of these
coefficients is made when evaluating Eq. (\ref{correlations}). 

Up to now, our model describes the coupling of the $n$ qubits of
the quantum computer to the environment. According to the above discussion,
we will make the assumption that only the subspace spanned by the
states $\{|m>,|p>\}$ is relevant for the dynamics. Accordingly, we
need to compute the matrix elements of Eq. (\ref{MODEL-V}) in this
basis. A straightforward calculation gives, in the limit of large
$N=2^{n}$:\begin{equation}
V=\sigma_{z}C,\label{Vred}\end{equation}

where $\sigma_{z}$ acts on the system subspace, and \begin{equation}
C=-\frac{1}{4}\sum_{i=1}^{n}(B_{i}+B_{i}^{\dagger})\end{equation}

acts on system E. In the interaction picture, the above operator becomes
\begin{equation}
C(t)=-\frac{n\lambda}{4}\sum_{n_{2}>n_{1}}c(n_{1},n_{2})e^{-it\omega(n_{1},n_{2})}|n_{1}\rangle\langle n_{2}|+h.c.,\label{Cinterpict}\end{equation}

with $\omega(n_{1},n_{2})=\frac{\delta\varepsilon}{N_{1}}(n_{2}-n_{1})$
. 

The interaction Hamiltonian Eq. (\ref{Vred}) is of the general form
(\ref{VprodS}), with only one term appearing. Consequently, only
one correlation function arises, which we represent by $G(s,s')$
. Associated to this function, it exists one function $\Gamma(t)$
defined as in (\ref{gammalk}). One can also check that condition
(\ref{trazacero}) is satisfied.

As for the state $\rho_{E}$ , we make the simplest choice, by taking
$\rho_{E}=\frac{I_{E}}{N_{1}}$ , where $I_{E}$ is the identity operator
in the environment space. This choice obviously satisfies $[H_{E},\rho_{E}]=0$
. As a consequence, the correlation function $G(s,s')$ only depends
on the difference $s-s'$ , and the function $\Gamma(t)$ becomes
independent of $t$ . A redefinition of the variables $s$ and $s'$
gives\begin{equation}
\Gamma=\lim_{\tau\rightarrow0}\frac{2}{\tau}\int_{0}^{\tau}ds\int_{0}^{s}ds'G(s').\label{limtau}\end{equation}

The correlation function can be calculated straightforwardly in the
present model. We will consider the limit $N_{1}\gg1$ . In this limit,
one obtains\begin{equation}
G(s)=N_{1}(\frac{n\lambda}{2\delta\varepsilon s}\sin\frac{\delta\varepsilon s}{2})^{2}.\end{equation}

The evaluation of the limit in Eq. (\ref{limtau}) deserves some discussion.
As will be shown below, we will be interested in time intervals which
are much larger than $1/\delta\varepsilon$ . For such long-time variations,
we can still consider values of $\tau$ that are larger than $1/\delta\varepsilon$,
i.e. we assume that $\tau\delta\varepsilon\gg1$ (see \cite{Michel}
for an extensive discussion). Within this context, Eq. (\ref{limtau})
finally gives \begin{equation}
\Gamma=\frac{n^{2}\lambda^{2}\pi N_{1}}{8\delta\varepsilon}.\end{equation}

The master equation (\ref{mastereq}) can be finally written, for
our model, as\begin{equation}
\frac{d}{dt}\rho_{S}(t)=-i[H_{S}(t),\rho_{S}(t)]+\Gamma(\sigma_{z}\rho_{S}(t)\sigma_{z}-\rho_{S}(t)).\label{masterGrov}\end{equation}

We have numerically solved Eq. (\ref{masterGrov}) using the model
presented above, for a system of $n=12$ qubits. We choose a value
of $\epsilon=0.1$, for which the Grover time is $t_{G}\simeq10³$
. The numerical values for the bath model are $N_{1}=2000$ , $\delta\varepsilon=0.5$
. Therefore $1/\delta\varepsilon=2\ll t_{G}$ , in agreement with
the approximations discussed previously. We compare our numerical
results to the solution of the Schrödinger equation of the total (S+E)
system. The initial state is $|\Psi(0)>=|\Psi_{S}(0)>\otimes|\Psi_{E}(0)>$
, where $|\Psi_{S}(0)>=\frac{1}{\sqrt{N}}|s>+\sqrt{\frac{N-1}{N}}|p>$
and $|\Psi_{E}(0)>=\frac{1}{\sqrt{N_{1}}}\sum_{n=1}^{N_{1}}|n>$ ,
consistent with the above choice of $\rho_{E}$ . 

According to the analysis based on the HAM equations, one expects
that Eq. (\ref{masterGrov}) will work when the conditions \begin{eqnarray}
c_{1}\equiv\frac{\lambda_{eff}N_{1}}{\delta\varepsilon} & \geq & \frac{1}{2}\nonumber \\
c_{2}\equiv\frac{\lambda_{eff}^{2}N_{1}}{\delta\varepsilon^{2}} & \ll & 1\label{criteria}\end{eqnarray}

are met \cite{Gemmer,Michel}, where the definition $\lambda_{eff}=\frac{1}{4}n\lambda$
arises as a consequence of Eq. (\ref{Cinterpict}). 

\begin{figure}
\includegraphics[width=8cm]{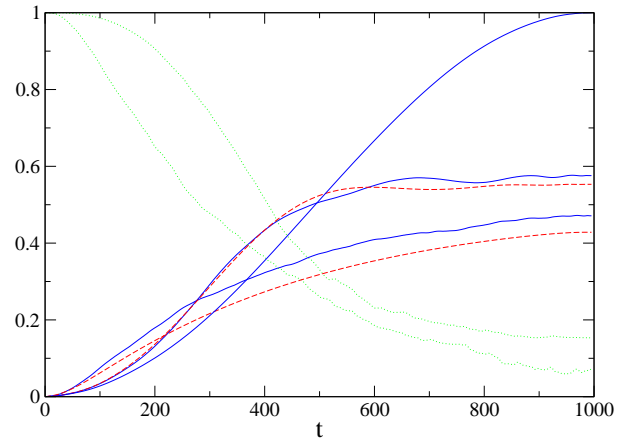}

\caption{(Color online) Numerical solution of the quantum search algorithm
using the exact Schrödinger equation (blue, solid curves) or the derived
master equation (\ref{masterGrov}) (red, dashed curves). We plot
the probability of overlapping with the searched state. Blue (solid)
curves are obtained, from top to bottom, for $\lambda=0$ , $\lambda=10^{-4}$
and $\lambda=5\times10^{-4}$. For the two last cases, the corresponding
approximated solution is also shown, also from top to bottom. Also
for these two cases, green (dotted) curves show how decoherence manifests
as time evolves, by plotting the magnitude defined in Eq. (\ref{cdeco}). }

\end{figure}

The results of our calculations are shown in Fig. 1. We plot the probability
of overlapping with the searched stated during the evolution of the
system. Solid (blue) lines are obtained from the exact solution to
the Schrödinger equation. The upper curve corresponds to the case
of no coupling to the environment ($\lambda=0)$ . Since the quantum
computer runs during a 'Grover time', the probability approaches unity.
The two lower curves have been obtained, from top to bottom, for $\lambda=10^{-4}$
and $\lambda=5\times10^{-4}$, giving $c_{1}=1.2,c_{2}=7.2\times10^{-4}$
and $c_{1}=6,c_{2}=1.8\times10^{-2}$, respectively. The interaction
with the environment translates into a worse performance of the adiabatic
search, which is manifested as a lower probability of success. This
effect becomes stronger as the coupling to the bath increases. The
degree of decoherence can be measured by several means. Here, as a
figure of merit we calculate the magnitude \cite{Montina}:\begin{equation}
\mathbb{C}=\sqrt{2Tr(\rho_{S}^{2})-1},\label{cdeco}\end{equation}

which is also shown in the same figure for the same values of $\lambda$
. Clearly, the decoherence increases with time. This effect is more
pronounced for a larger coupling, giving rise to an almost completely
incoherent, and equally probable mixture, of the $\{|m>,|p>\}$ states.

One can also observe that the approximation obtained by solving the
derived master equation (red, dashed curves) becomes more accurate
for lower values of the coupling, in accordance to criteria Eq. (\ref{criteria}).
Indeed, most of the difference observed for the smallest $\lambda$
are due to oscillations , corresponding to the fact that the complete
numerical solution has been obtained for a particular realization
of the couplings in Eq. (\ref{couplings}), while these constants
have been averaged out in obtaining (\ref{masterGrov}).

\section{Conclusions}

In this paper, we have first discussed the so-called Hilbert space
Average Method, as an alternative to describe open quantum systems.
We extended the method to the case of a system subject to a time-dependent
Hamiltonian. We also made a connection of the evolution equations
for this method with known master equations. 

We next discussed a simple model which can be useful for the study
of a quantum computer performing an adiabatic quantum search, while
in contact with an environment. The ultimate purpose of such study
is, of course, the understanding of the effects of decoherence on
the performance of the computation. The model for the environment
is simply a band of equally spaced levels with random coupling to
the qubits of the quantum computer. In spite of its simplicity, we
have shown that it incorporates decoherence effects in a clear way.
The equations for the reduced system can be studied either under the
form of HAM dynamics or master equations.

One can also, for this model, perform an exact numerical simulation
of the total system (including the environment). We have performed
such a numerical study, and compared the results with the approximated
dynamics of the system. As expected, increasing the strength of the
coupling between the system and the environment implies a larger degree
of decoherence, which translates into a lower probability of success
for the quantum search. On the other hand, increasing the coupling
also means that the master equation gives a poorer description of
the actual dynamics. The degree of approximation is controlled by
the criteria derived for HAM equations within similar models for the
environment. 

\textbf{Acknowledgments}

I would like to acknowledge the comments made by M.C. Bañuls, I. de
Vega and A. Romanelli, during interesting discussions, and also the
hospitality of the Max-Planck-Institut für Quantenoptik in Munich.
This work has been supported by the Spanish Ministerio de Educación
y Ciencia through Projects AYA2007-67626-C03-C1 and FPA2005-00711.

\end{document}